\documentclass[aps,prb,twocolumn,showpacs,groupedaddress,preprintnumbers,amsmath,amssymb]{revtex4}

\usepackage{graphicx}
\usepackage{amsmath,amssymb,epsfig,color}
\usepackage{dcolumn}
\usepackage{bm}

\begin{document}

\title{Dynamical magnetic susceptibility in the lamellar cobaltate superconductor Na$_x$CoO$_2$$\cdot y$H$_2$O}

\author {M.M.~Korshunov $^{1,2}$}
 \email {maxim@mpipks-dresden.mpg.de}
\author {I.~Eremin $^{1,3}$}
 \affiliation {$^1$ Max-Planck-Institut f\"{u}r Physik komplexer Systeme, D-01187 Dresden, Germany}
 \affiliation {$^2$ L.V. Kirensky Institute of Physics, Siberian Branch of Russian Academy of Sciences, 660036 Krasnoyarsk, Russia}
 \affiliation {$^3$ Institute f\"{u}r Mathematische und Theoretische Physik, TU Braunschweig, 38106 Braunschweig, Germany}

\date{\today}

\begin{abstract}
We systematically analyze the influence of the superconducting gap
symmetry and the electronic structure on the dynamical spin
susceptibility in superconducting Na$_x$CoO$_2$$\cdot y$H$_2$O
within a three different models: the single $a_{1g}$-band model
with nearest-neighbor hoppings, the realistic three-band
$t_{2g}$-model with, and without $e'_g$ pockets present at the Fermi
surface. We show that the magnetic response in the normal state is
dominated by the incommensurate antiferromagnetic spin density wave
fluctuations at large momenta in agreement with experimental
temperature dependence of the spin-lattice relaxation rate. Also, we
demonstrate that the presence or the absence of the
$e'_g$-pockets at the Fermi surface does not affect significantly
this conclusion. In the superconducting state our results for
$d_{x^2-y^2}$- or $d_{xy}$-wave symmetries of the superconducting
order parameter are consistent with experimental data and exclude
nodeless $d_{x^2-y^2} + id_{xy}$-wave symmetry. We further point out
that the spin-resonance peak proposed earlier is improbable for the
realistic band structure of Na$_x$CoO$_2$$\cdot y$H$_2$O. Moreover,
even if present the resonance peak is confined to the antiferromagnetic
wave vector and disappears away from it.
\end{abstract}


\maketitle

\section{Introduction}

The spin dynamics in unconventional non $s$-wave superconductors is
of fundamental interest due to its interesting and peculiar
properties. This includes a non-trivial behavior of the magnetic
part of the Knight shift in the spin-triplet superconductors
\cite{woellfle}, as well as an emergence of the so-called resonance
peak observed in superconducting layered cuprates \cite{bourges}
which possesses spin-singlet $d_{x^2-y^2}$-wave order parameter
symmetry. Furthermore, magnetic excitations are also often
considered as a possible glue for the Cooper-pairing in a number of
heavy-fermion and transition metal oxides compounds.

An analysis of the feedback effect of superconductivity on the magnetic spin
susceptibility can be used to determine the symmetry of the superconducting
order parameter. This is of particular significance for recently discovered
water intercalated sodium cobaltate superconductor \cite{Takada2003},
Na$_x$CoO$_2 \cdot y$H$_2$O, where the origin of superconductivity as well as
an underlying symmetry of the superconducting order parameter is currently
under debate. The studies of the specific heat
\cite{Cao2003,Lorenz2004,Yang2005,Jin2005} and the $\mu$SR measurements of a
magnetic penetration depth \cite{Kanigel2004} have revealed a line of nodes in
the superconducting gap function $\Delta_{\bf k}$. Similar conclusion has been
made based on the measurements of the spin-lattice relaxation rate $1/T_1T$ by
means of Nuclear Quadrupole Resonance (NQR), where absence of the
characteristic Hebel-Slichter peak and power-law decrease upon decreasing
temperature has been observed
\cite{Fujimoto2004,Ishida2003,Zheng2006,Ihara2006,Michioka2006}.
Simultaneously, the developing of the strong antiferromagnetic (AFM)
fluctuations above superconducting transition temperature, $T_c$, have been
found. At the same time, early reports on the Knight shift's temperature
dependence, $K(T)$, have suggested a spin-triplet symmetry of the
superconducting gap \cite{Kato2005,Ihara2006_2}. In these Nuclear Magnetic
Resonance (NMR) experiments, $K(T)$ was shown to be anisotropic for external
magnetic field applied parallel or perpendicular to the $ab$-plane. In
particular, $K_c(T)$ component has not shown a substantial decrease below
$T_c$. This behavior has been interpreted in favor of the odd-parity
Cooper-pairing in sodium cobaltates
\cite{Tanaka2003,Motrunich2004,Kuroki2004,Johannes2004,Kuroki2005,Mochizuki2005}.
However, the most recent NMR experiments with higher precision have found a
reduction of both Knight shift components as a function of temperature for
$T<T_c$ \cite{Zheng2006_2,Kobayashi2006}. These experiments points towards
spin-singlet Cooper-pairing.

From the group-theoretical analysis the even-parity symmetries of
the lowest harmonics for the triangular lattice are classified
according to $s$-wave ($\Delta_{\bf k}=\Delta_0$), extended-$s$-wave
($\Delta_{\bf k}=2/3 \Delta_0 [\cos{k_y} + 2 \cos{(k_x \sqrt{3}/2)} \cos{(k_y/2)}]$),
$d_{x^2-y^2}$-wave
($\Delta_{\bf k}=\Delta_0 [\cos{k_y} - \cos{(k_x \sqrt{3}/2)} \cos{(k_y/2)}]$),
$d_{xy}$-wave
($\Delta_{\bf k}=\Delta_0 [\sqrt{3} \sin{(k_x \sqrt{3}/2)} \sin{(k_y/2)}]$),
and $d_{x^2-y^2}+id_{xy}$-wave representations \cite{Mazin2005}.
For both $d_{x^2-y^2}$-wave and $d_{xy}$-wave
symmetries $\Delta_{\bf k}$ has line of nodes at the Fermi surface.
Moreover, the time-reversal symmetry is broken for
$d_{x^2-y^2}+id_{xy}$-wave state.

For the pure trigonal symmetry of the CoO$_2$-plane, all three $d$-wave states
are degenerate. However, due to the absence of nodes $d_{x^2-y^2}+id_{xy}$-wave
seems to be most energetically favorable. Until now, a breaking of
time-reversal symmetry has not been observed in experiment
\cite{Higemoto2004,Higemoto2006}. Generally, the combined influence of the
impurities and some competing instabilities, such as Cooper-pairing in a
secondary channel as well as the lattice symmetry breaking, can lift the
degeneracy between these three $d$-wave competing ground states
\cite{Florens2005}. This may indeed be the case for sodium cobaltates where Na
arrangement introduces disorder at $x=0.33$ concentration
\cite{Zandbergen2004}. More sophisticated theories, involving multi-orbital
model for sodium cobaltates, suggest two different gap symmetries (one of which
is $d_{x^2-y^2}+id_{xy}$) for two different Fermi surface topologies
\cite{Mochizuki2007}.

Obviously, there is still a controversy on the symmetry of the
superconducting order parameter in sodium cobaltates. In present
study we systematically analyze the influence of the superconducting
(SC) gap symmetry and the electronic structure on the dynamical spin
susceptibility in Na$_x$CoO$_2$$\cdot y$H$_2$O. In particular,
assuming spin singlet $s$-wave and $d$-wave symmetries of the
superconducting order parameter we have calculated the real and the
imaginary part of the magnetic response as a function of the
momentum, temperature and frequency. We deduce the characteristic
temperature dependencies of the Knight shift and spin-lattice
relaxation rate. Furthermore, we have studied the feedback of the
superconducting order parameter on the frequency dependence of the
imaginary part of the spin susceptibility. We investigate the role
played by the details of the electronic structure of
Na$_x$CoO$_2$$\cdot y$H$_2$O and, in particular, the changes of the
Fermi surface (FS) topology induced by the multi-orbital effects.

Structurally, a parent compound, Na$_x$CoO$_2$, has a
quasi-two-dimensional structure with Co ions in the CoO$_2$ layers
forming a triangular lattice. Na ions reside between these layers
and donate $x$ electrons to the partially filled Co-$d(t_{2g})$
orbital. Apart from doping, Na ions also induce structural ordering
at higher doping concentrations ($x \geq 0.5$) where
superconductivity does not occur. Due to the presence of a trigonal
crystalline electric field (CEF), the $t_{2g}$ level splits into the
higher lying $a_{1g}$ singlet and the two lower lying $e'_g$ states.
The {\it ab-initio} band structure calculations within a Local
Density Approximation (LDA) predict Na$_x$CoO$_2$ to have a large
Fermi surface with mainly $a_{1g}$ character and six hole pockets of
mostly $e'_g$ character \cite{djs2000}. At the same time, surface
sensitive Angle-Resolved Photo-Emission Spectroscopy (ARPES)
\cite{mzh2004, hby2004, shimojima2006} reveals a doping dependent
evolution of the Fermi surface, which shows no sign of the $e'_g$
hole pockets for $0.3 \le x \le 0.8$. Instead, the observed Fermi
surface is centered around the $\Gamma$ point and has mostly
$a_{1g}$ character. It has been argued that such an effect may arise
due to strong electronic correlations
\cite{Zhou2005,Korshunov2007,Shorikov2007}, however, no consensus in
the literature has been reached yet (see e.g.
\cite{Ishida2005,Perroni2007,Liebsch2007}).

In Na$_x$CoO$_2$$\cdot y$H$_2$O due to the water
intercalation the inter-layer CoO$_2$ distance becomes larger and,
thus, the material becomes more two-dimensional leading to a
substantial decrease of the bilayer splitting. However, little is
known about the particular changes in the electronic structure and
the energy splitting between $a_{1g}$ and $e'_g$ levels.

In order to take into account the multi-orbital effects we analyze
the effect of superconductivity for the three different cases: the
single-band ($a_{1g}$) model with nearest-neighbor hoppings, the
realistic three-band ($t_{2g}$) model with, and without six $e'_g$
pockets at the FS.

\section{$a_{1g}$-band model}

\begin{figure}
\includegraphics[angle=0,width=1.0\linewidth]{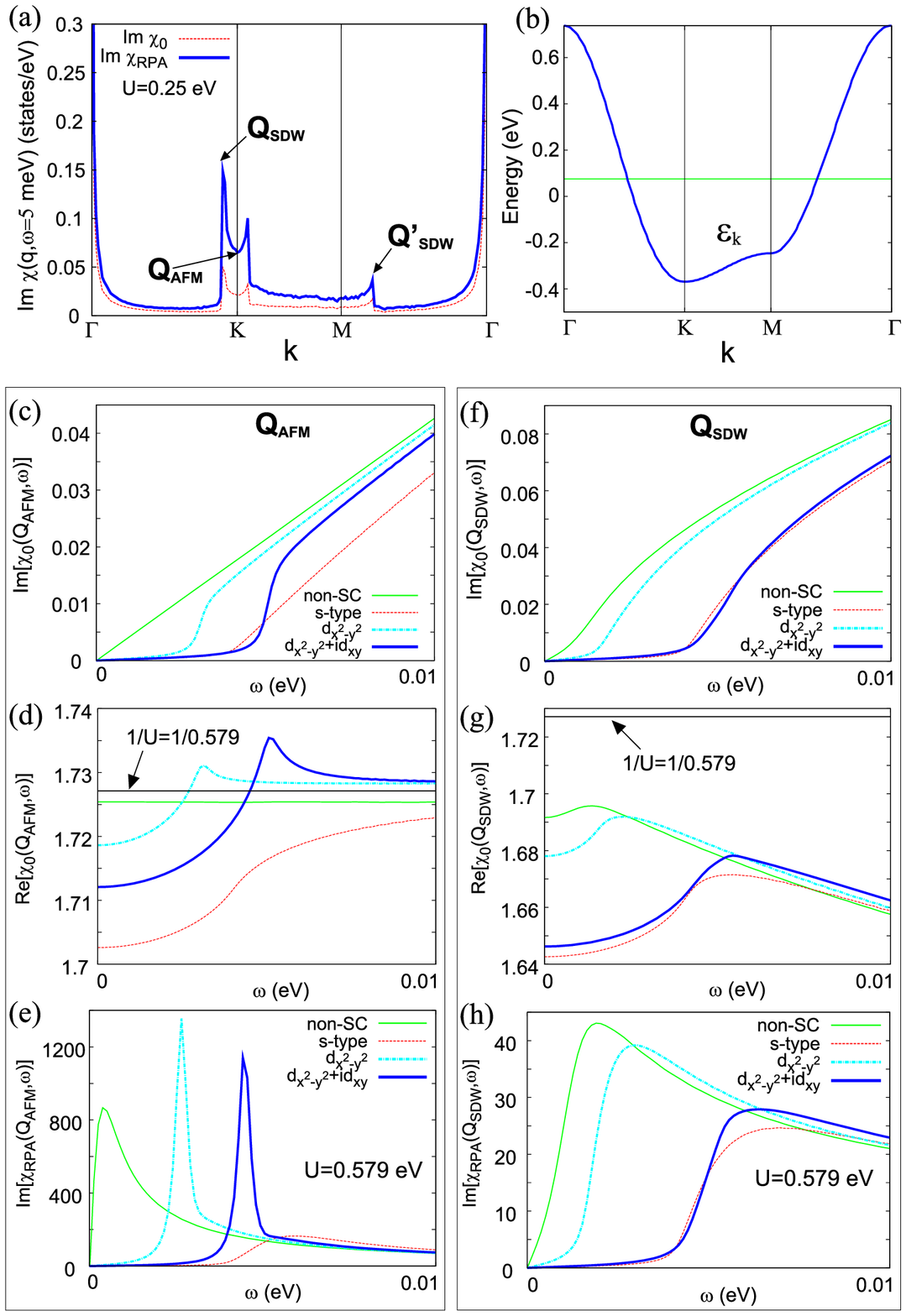}
\caption{(Color online) Calculated results for the $a_{1g}$-band
model. (a) ${\bf q}$-dependence of ${\rm Im}[\chi_0({\bf
q},\omega)]$ and ${\rm Im}[\chi_{RPA}({\bf q},\omega)]$ at
$\omega=5$ meV in the normal (non-SC) phase. The scattering wave
vectors ${\bf Q_{AFM}}$, ${\bf Q_{SDW}}$, and ${\bf Q'_{SDW}}$ are
denoted by the arrows.
(b) Calculated $a_{1g}$-band dispersion,
where the horizontal (green) line stands for the chemical
potential.
The panels (c)-(e) show imaginary and real
parts of $\chi_{0}$, and imaginary part of $\chi_{RPA}$ at ${\bf q=Q_{AFM}}$
in non-SC phase and in SC phase with various superconducting
order parameter symmetries. The same quantities are plotted
in the panels (f)-(h) at the wave vector ${\bf q=Q_{SDW}}$.
Here we choose the amplitude of the superconducting order parameter
$\Delta_0=2$ meV. For the numerical purposes we also employ the
broadening of the Green's function, $\delta=0.2$ meV.}
\label{fig:a1g_chi}
\end{figure}

\begin{figure}
\includegraphics[angle=0,width=0.7\linewidth]{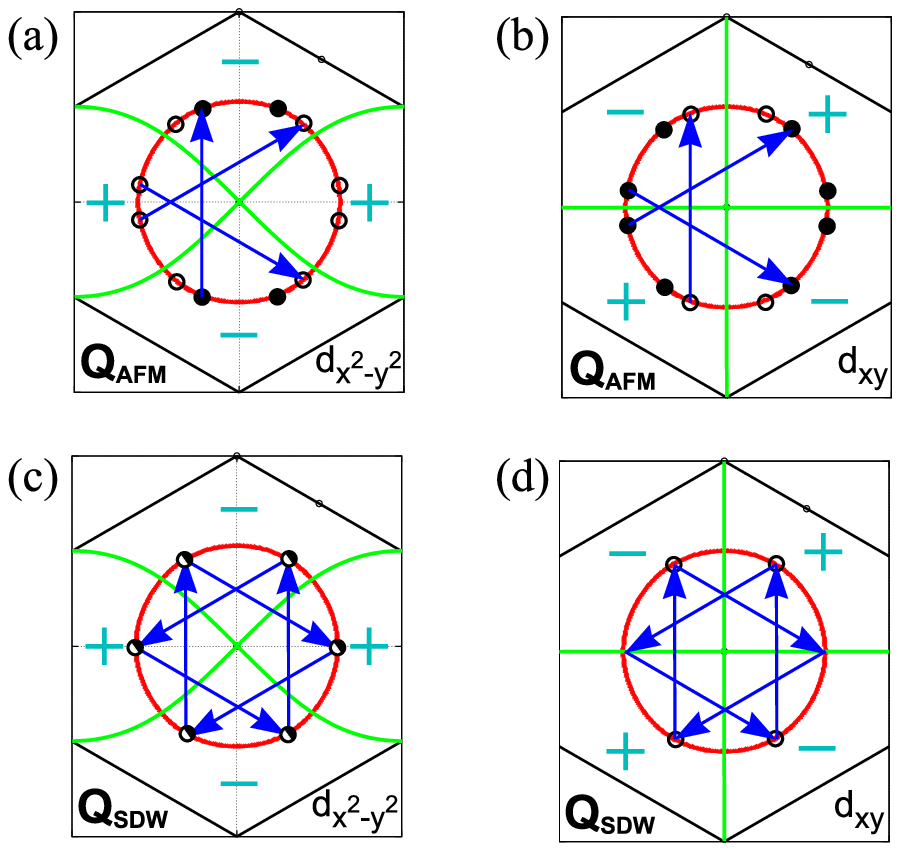}
\caption{(Color online) Calculated Fermi surface
of the $a_{1g}$-band model. The position of the nodes of the
$d_{x^2-y^2}$-wave [(a),(c)] and $d_{xy}$-wave [(b),(d)]
superconducting gaps is denoted by the solid (green) curves. The
plus and the minus signs refer to the corresponding phases of the
superconducting order parameter. The states at the FS connected by
the wave vectors ${\bf Q_{AFM}}$ or ${\bf Q_{SDW}}$ are shown by the
circles.} \label{fig:a1g_SCgap}
\end{figure}

We first consider the simple $a_{1g}$-band model, represented by a
two-dimensional Hubbard Hamiltonian on the triangular lattice:
\begin{equation}
H = -\sum\limits_{{\bf k}, \sigma} \varepsilon_{{\bf k}} a_{{\bf k} \sigma}^\dag a_{{\bf k} \sigma}
+ \sum\limits_{i } U n_{i \uparrow} n_{i \downarrow},
\label{eq:Hint}
\end{equation}
where $n_{i \sigma} = a_{i  \sigma}^\dag a_{i \sigma}$, $a_{i \sigma}$ ($a_{i
\sigma}^\dag$) is the annihilation (creation) operator for the $a_{1g}$ hole at
the Co site $i$ with spin $\sigma$. Here, $\varepsilon_{\bf k} = 2 t [\cos{k_y}
+ 2 \cos{(k_x \sqrt{3}/2)} \cos{(k_y/2)}] - \mu$, $t$=0.123 eV is the
nearest-neighbor hopping integral, and $\mu$ is the chemical potential which
has been calculated self-consistently for $x=0.33$. The energy dispersion,
$\varepsilon_{\bf k}$,  along the principal directions of the hexagonal
Brillouin zone (BZ) and the corresponding Fermi surface are shown in
Fig.~\ref{fig:a1g_chi}(b) and in Fig.~\ref{fig:a1g_SCgap}, respectively. Here,
$\Gamma=(0,0)$, ${\rm K}=(0,2/3)$, and ${\rm M}=(1/2\sqrt{3},1/2)$ [in units of
$2\pi/a$] denote the symmetry points of the first BZ. Later, coordinates of the
wave vectors will be given in units of $2\pi/a$ with $a$ being the in-plane
lattice constant.

To calculate the dynamical spin susceptibility, we employ the Random
Phase Approximation (RPA) which gives
\begin{equation}
\chi_{RPA} ({\bf q}, i \omega_m) =  \frac{\chi_0 ({\bf q}, i
\omega_m)}{1-U \chi_0 ({\bf q}, i \omega_m)},
\end{equation}
where $\chi_0({\bf q}, i \omega_m)$ is the BCS Lindhard
susceptibility
\begin{eqnarray}
\chi_0 ({\bf q}, i \omega_m) =  \frac{1}{2N}\sum\limits_{{\bf k}} \left[
\frac{ f(E_{{\bf k+q}}) - f(E_{{\bf k}})}{i\omega_m - E_{{\bf k+q}} + E_{{\bf k}}} C_{{\bf k},{\bf q}}^{+} \right. \nonumber\\
+ \frac{1 - f(E_{{\bf k+q}}) - f(E_{{\bf k}})}{2} C_{{\bf k},{\bf q}}^{-} \nonumber\\
\times \left.
\left( \frac{1}{i\omega_m + E_{{\bf k+q}} + E_{{\bf k}}} - \frac{1}{i\omega_m - E_{{\bf k+q}} - E_{{\bf k}}}
\right) \right],
\end{eqnarray}
with $C_{{\bf k},{\bf q}}^{\pm} = 1 \pm \frac{\varepsilon_{\bf k}
\varepsilon_{\bf k+q} + {\rm Re}{(\Delta_{\bf k} \Delta_{\bf k+q}^*)}} {E_{{\bf
k}} E_{{\bf k+q}}}$ being the BCS coherence factors. Here, $\omega_m$ are the
Matsubara frequencies, $f(E)$ is the Fermi function, and $E_{{\bf
k}}=\sqrt{\varepsilon_{\bf k}^2 + |\Delta_{\bf k}|^2}$.

In Fig.~\ref{fig:a1g_chi}(a) we show both the bare and the RPA
magnetic susceptibility in the normal state at $\omega=5$ meV and
$U=0.25$ eV. One immediately notices that the magnetic response is
dominated by the scattering at the {\it incommensurate} wave vector,
${\bf Q_{SDW}}=(0,0.598) \approx (0,3/5)$.
The value of ${\rm Im}[\chi({\bf q},\omega)]$ at the commensurate wave
vector,
${\bf Q_{AFM}}=\left\{(0,2/3), (1/\sqrt{3},1/3)\right\}$,
appears to be much smaller. There is
also another incommensurate wave vector present, ${\bf Q'_{SDW}}$.
The presence of a set of incommensurate wave vectors with substantial
magnitude of magnetic scattering shows a tendency of the itinerant
electrons on the triangular lattice towards spin density wave (SDW)
instability.

In Fig.~\ref{fig:a1g_chi}(c) and (d) we present the imaginary and the real
parts of $\chi_{0}({\bf Q_{AFM}},\omega)$ as a function of frequency $\omega$
at $T=1$ K. In the non-SC state, the imaginary part is linear in $\omega$ at
low frequencies which is a typical Landau damping within the Fermi-liquid
picture. In the SC phase, the imaginary part of the magnetic susceptibility
becomes gapped. The magnitude of the gap, $\Omega_g$, is equal to $2\Delta_0$
in the $s$-wave case. At larger frequencies ${\rm Im}\chi_{0}$ increases slowly
from zero. In comparison, for the $d$-wave symmetries the lowest value of
$\Omega_g = |\Delta_{\bf k}| + |\Delta_{\bf k+Q}|$ at the Fermi surface.
Obviously for the non-$s$-wave symmetry it is smaller than $2\Delta_0$. From
Fig.~\ref{fig:a1g_SCgap}(a) and (b) one can notice that the ${\bf Q_{AFM}}$
wave vector connects parts of the FS where $\Delta_{\bf k} = - \Delta_{\bf
k+Q_{AFM}}$ but also some parts where $\Delta_{\bf k} = + \Delta_{\bf
k+Q_{AFM}}$. For $d_{x^2-y^2}$-wave superconducting gap, there are four pairs
of points of the first type and two pairs of points of the second type. Due to
the smaller $|\Delta_{\bf k}| + |\Delta_{\bf k+Q_{AFM}}|$ for the first
process, as it is seen from Fig.~\ref{fig:a1g_SCgap}(a), the ${\rm Im}\chi_0$
shows a discontinuous jump at $\Omega_g$. This is due to the change of sign in
the anomalous coherence factor, $C_{\bf k,q}^{-}$. The second process will give
contribution at energies larger than $\Omega_g$ due to larger value of
$|\Delta_{\bf k}| + |\Delta_{\bf k+Q_{AFM}}|$ there.  Therefore, the net effect
will result in a discontinuous jump of ${\rm Im}\chi_0$ at $\Omega_g$.
Correspondingly, the real part will possess a logarithmic singularity as it is
also seen in Fig.~\ref{fig:a1g_chi}(d). Within the RPA the formation of the
pole (spin resonance) in the total magnetic susceptibility below $\Omega_g$ is
possible if ${\rm Im}[\chi_{0}({\bf q},\omega)]=0$ and simultaneously $1/U={\rm
Re}[\chi_{0}({\bf q},\omega)]$. Due to the logarithmic character of the
singularity this condition will be generally fulfilled for any small value of
$U$ which would give a position of the resonance exactly at or very close to
$\Omega_g$. However, a small amount of impurities or disorder will smear the
singularity out and suppress the resonance peak. In Na$_x$CoO$_2$$\cdot
y$H$_2$O the value of $U$ should be relatively large to shift the position of
the spin resonance towards energies smaller than $\Omega_g$ and make it robust
against impurity scattering. The calculated susceptibility is shown in
Fig.~\ref{fig:a1g_chi}(e) where we use $U_{res}$=0.579 eV. It is interesting to
note that the resonance occurs for both $d_{x^2-y^2}$- and
$d_{x^2-y^2}+id_{xy}$-wave symmetries, however, the value of $\Omega_g$
slightly differs. Note, for $d_{xy}$-wave superconducting gap the situation is
opposite. From Fig.~\ref{fig:a1g_SCgap}(b) one sees that in contrast to
$d_{x^2-y^2}$-wave case there are two pairs of points at the FS where
$\Delta_{\bf k} = - \Delta_{\bf k+Q_{AFM}}$ and four pairs of points where
$\Delta_{\bf k} = + \Delta_{\bf k+Q_{AFM}}$. Here, the $\Omega_g$ is determined
by the second process, thus there will be no logarithmic jump in ${\rm
Re}\chi_0$ at $\Omega_g$. Of course it will occur at larger frequencies due to
the first type of process but the resonance conditions will not be fulfilled.
Therefore, we do not expect the spin resonance for the $d_{xy}$-wave symmetry.

The present value of $U_{res}$ is of course too small to be the
on-site Coulomb repulsion which is of the order of several electron
volts. Therefore, the effective interaction $U$ entering our model
(\ref{eq:Hint}) originates mainly from the Hund's exchange, $J_H$.
In the lamellar sodium cobaltate, the value of $J_H$
is presently disputed and the lowest estimated value is of the order
of 1eV \cite{Shorikov2007}. It has been shown recently that even
this value significantly affects the population of the $a_{1g}$ and
$e'_g$ orbitals\cite{Ishida2005,Perroni2007,Liebsch2007}. Taking
this value into account, we assume $U=\alpha J_H$, where $J_H$ is
the mean-field value of the Hund's exchange and $\alpha$ is the
coefficient that describes corrections beyond mean-field theory. One
has to note that the larger value of $U$ will lead to the SDW
instability in our calculations.
\begin{figure}
\includegraphics[angle=0,width=1.0\linewidth]{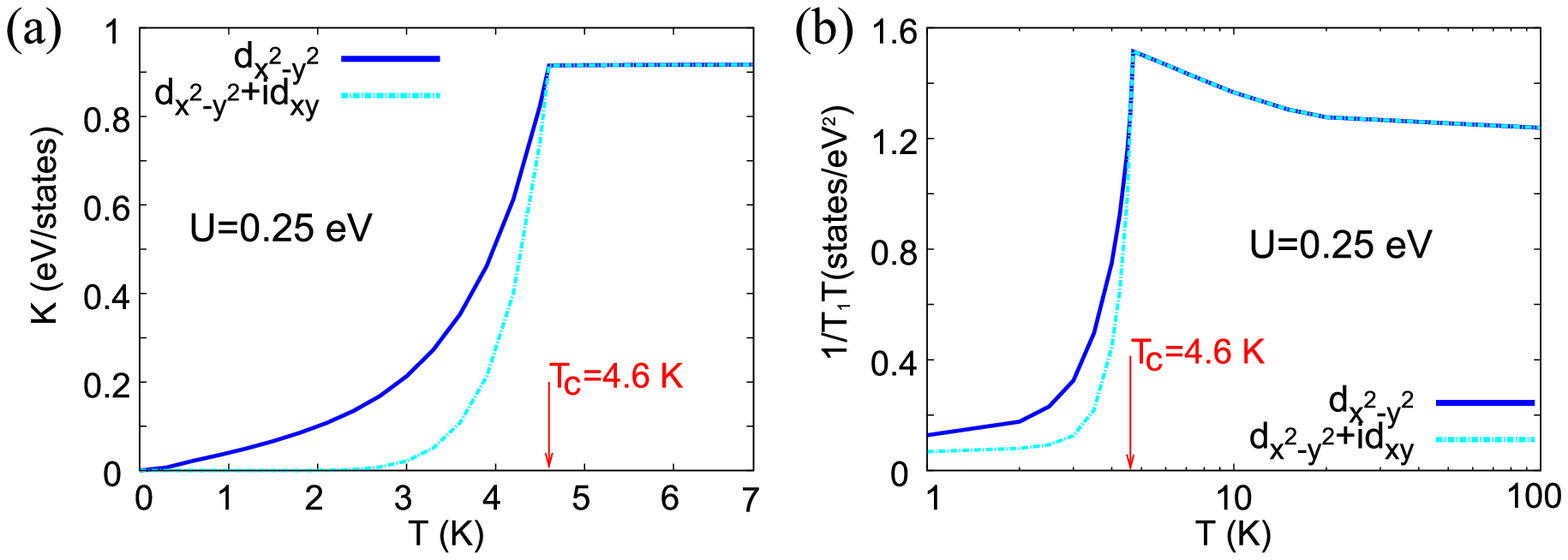}
\caption{Calculated  temperature dependence of the Knight shift
$K(T)$ (a) and the spin-lattice relaxation rate $1/T_1T$ (b) for the
$a_{1g}$-band model. Note the logarithmic temperature scale in (b).
Here, we assume the conventional BCS temperature dependence of
superconducting gap, $\Delta_0(T)=\Delta_0 \sqrt{1-T/T_c}$.}
\label{fig:a1g_nn_K_1_T1T}
\end{figure}

The situation changes for the wave vector ${\bf Q_{SDW}}$
[Fig.~\ref{fig:a1g_chi}(f)-(h)]. There is one striking difference in
the low-energy behavior of ${\rm Im}[\chi_{0}({\bf
Q_{SDW}},\omega)]$. Namely, already in the normal state the
scattering rate is {\it non-linear} for small $\omega$. It is
obviously a consequence of the $2{\bf k_F}$ instability and a
resulting non-Landau damping at this wave vector. Furthermore, in
the SC state the situation differs drastically with respect to ${\bf
Q_{AFM}}$. As one could see from Fig.~\ref{fig:a1g_SCgap}(c) and (d)
there is equal number of contributions for which $\Delta_{\bf k} = -
\Delta_{\bf k+Q_{SDW}}$ and $\Delta_{\bf k} = + \Delta_{\bf
k+Q_{SDW}}$. As a result the discontinuity does not occur and the
real part of $\chi_0$ is smaller in the superconducting state than
in the normal state. Therefore, for reasonable values of $U$ there
is no resonance condition for $\chi_{RPA}$ [see
Fig.~\ref{fig:a1g_chi}(h)].

Generally, a formation of the resonance peak below $T_c$ in the unconventional
superconductors is a well-known consequence of the sign change of the
superconducting order parameter. It has been originally discussed in relation
to the layered high-$T_c$ cuprates \cite{high-Tc_Resonance} and also recently
has been used to explain the inelastic neutron scattering results in
heavy-fermion compound UPd$_2$Al$_3$ \cite{Chang2007}. In layered
superconducting cobaltates the emergence of the resonance peak for several
symmetries of the superconducting order parameter has been analyzed within
simple single-band model \cite{Li2004}. In contrast to
Ref.~\onlinecite{Li2004}, we have found that the resonance peak (even within
simple $a_{1g}$-band model) is very sensitive to the small variation of
$U$-values and to disorder. As a result the resonance is confined to the wave
vector ${\bf Q_{AFM}}$ and disappears for $\left| {\bf Q} \right| < \left| {\bf
Q_{AFM}} \right|$.

The temperature dependence of the Knight shift, $K(T)$, and the
spin-lattice relaxation rate, $1/T_1T$, is calculated according to
the expressions:
\begin{eqnarray}
K(T) & \propto & \lim\limits_{{\bf q} \rightarrow 0} {\rm Re} \chi({\bf q}, \omega=0), \\
1/T_1T & \propto & \lim\limits_{\omega \rightarrow 0} \frac{1}{\pi}
\sum\limits_{\bf q}\frac{{\rm Im} \chi({\bf q}, \omega)}{\omega}.
\end{eqnarray}
In Fig.~\ref{fig:a1g_nn_K_1_T1T} we show both quantities as a
function of temperature. In the normal state $1/T_1 T$ increases
with decreasing temperature that reflects the presence of the
incommensurate antiferromagnetic fluctuations in this system. At the
same time, the Knight shift is a constant which stresses that there
are no small-{\bf q} fluctuations. Below $T_c$ both physical
observables drop rapidly due to opening of the superconducting gap
in the energy spectrum. As expected, the decrease is exponential for
$d_{x^2-y^2}+id_{xy}$-wave symmetry due to its nodeless character in
$E_{\bf k}$. For $d_{x^2-y^2}$-wave symmetry the behavior of $1/T_1 T$
and $K(T)$ follows standard power-law temperature dependence due to
the presence of the line nodes in the energy spectrum. In the next
section we will compare our results to the experimental data
where we describe a more realistic model in application to the
superconducting cobaltate.

\section{$t_{2g}$-band model}

The $a_{1g}$-band model is, of course, oversimplified for describing
the physics of Na$_x$CoO$_2$$\cdot y$H$_2$O since $a_{1g}$-$e'_g$
level splitting, $\delta\epsilon$, is only 53 meV. As a result
there is a substantial hybridization of the $a_{1g}$ and the $e'_g$
bands, completely neglected in the simple $a_{1g}$-band model. In
particular, the $e'_g$ bands may form hole pockets at the FS
in addition to a large $a_{1g}$-pocket \cite{djs2000}. To take into
account these details, we further analyze the magnetic response in
the full $t_{2g}$-band model including both $a_{1g}$ and $e'_g$
cobalt states.
\begin{figure}[htb]
\includegraphics[angle=0,width=0.8\linewidth]{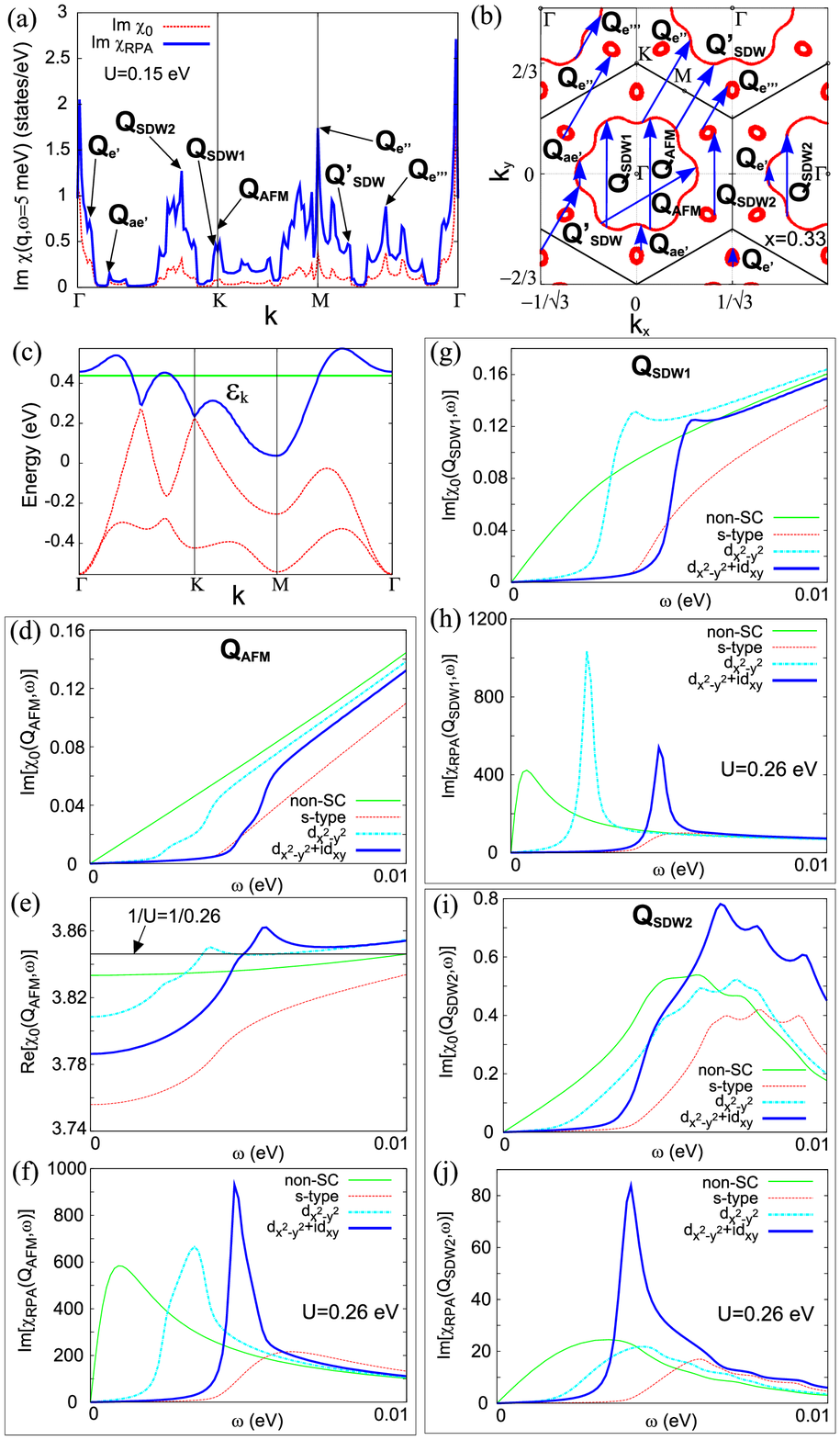}
\caption{(Color online) Calculated results for the $t_{2g}$-band
model. (a) ${\bf q}$-dependence of ${\rm Im}[\chi_0({\bf
q},\omega)]$ and ${\rm Im}[\chi_{RPA}({\bf q},\omega)]$ at
$\omega=5$ meV in the normal (non-SC) phase. The scattering wave
vectors ${\bf Q_{AFM}}$, ${\bf Q_{SDW1}}$, ${\bf Q_{SDW2}}$, ${\bf
Q'_{SDW}}$, ${\bf Q_{ae'}}$, ${\bf Q_{e'}}$, ${\bf Q_{e''}}$, and
${\bf Q_{e'''}}$  are denoted by the arrows. (b) The calculated
Fermi surface with the corresponding scattering wave vectors. In (c)
the band dispersion is shown where the bold (blue) curve denotes the
topmost band used for the susceptibility calculations. A horizontal
(green) line stands for the chemical potential.  The panels (d)-(f)
show imaginary and real parts of $\chi_{0}$, and imaginary part of
$\chi_{RPA}$ at ${\bf q=Q_{AFM}}$ in the normal state and in SC
state with various superconducting order parameter symmetries. The
imaginary parts of the bare and the total susceptibilities are
plotted in the panels (g)-(h) and (i)-(j) at the wave vectors ${\bf
q=Q_{SDW1}}$ and ${\bf q=Q_{SDW2}}$, respectively. Here we choose
the amplitude of the superconducting order parameter $\Delta_0=2$
meV. For the numerical purposes we also employ the broadening of the
Green's function, $\delta=0.2$ meV.} \label{fig:rigidband_chi}
\end{figure}

The free electron Hamiltonian of the $t_{2g}$-band model in a hole
representation is given by
\begin{equation}
H_0 = - \sum\limits_{{\bf k},\alpha ,\sigma } {\left( {\epsilon
^\alpha - \mu } \right)n_{{\bf k} \alpha \sigma } } -
\sum\limits_{{\bf k}, \sigma} \sum\limits_{\alpha, \beta}
t_{{\bf k}}^{\alpha \beta } d_{{\bf k} \alpha \sigma }^\dag d_{{\bf k} \beta \sigma},
\label{eq:H0}
\end{equation}
where $n_{{\bf k} \alpha \sigma} = d_{{\bf k} \alpha \sigma}^\dag
d_{{\bf k} \alpha \sigma}$, $d_{{\bf k} \alpha \sigma}$ ($d_{{\bf k}
\alpha \sigma}^\dag$) is the annihilation (creation) operator for
the $t_{2g}$-hole with spin $\sigma$, orbital index $\alpha$, and
momentum ${\bf k}$, $t_{{\bf k}}^{\alpha \beta}$ is the hopping
matrix element, $\epsilon^{\alpha}$ is the single-electron energy,
and $\mu$ is the chemical potential. All of the in-plane hoppings
and the single-electron energies were derived previously by us from
the {\it ab-initio} LDA calculations using projection procedure and
we use here the parameters for $x$=0.33 from
Ref.~\onlinecite{Korshunov2007}. To obtain the dispersion we
diagonalize the Hamiltonian (\ref{eq:H0}) calculating the chemical
potential $\mu$ self-consistently. The resulting FS topology and
energy dispersion are shown in Fig.~\ref{fig:rigidband_chi}(b) and
(c), respectively. The resulting dispersion and the FS replicate the
corresponding LDA ones\cite{Korshunov2007}.

Due to the non-zero inter-orbital hopping matrix elements, $a_{1g}$
and $e'_{g}$ bands are hybridized. However, only one of the
hybridized bands crosses the Fermi level thus making the largest
contribution to the low-energy properties of the system. We refer to
this band as $\varepsilon_{\bf k}$. Note, it is substantially
different from the simple $a_{1g}$-band. Later, this effective band $\varepsilon_{\bf k}$
will be used to calculate the dynamical magnetic susceptibility with
some effective on-site Coulomb interaction $U$.

Present FS has more complicated structure in comparison to the $a_{1g}$-band
model. First, $e'_g$ states are present at the Fermi surface and strongly
hybridize with $a_{1g}$ states. At the same time, the ``rounded hexagon'' shape
of the central part of the FS arises from the hoppings beyond nearest-
neighbors included in the $t_{2g}$-band model and neglected in $a_{1g}$-band
model considered above. This results in a number of additional scattering wave
vectors as calculated from $\chi_0$, see Fig.~\ref{fig:rigidband_chi}(a). In
particular, there are four scattering wave vectors connecting the $e'_g$-$e'_g$
FS pockets [${\bf Q_{e'}}$, ${\bf Q_{e''}}$, ${\bf Q_{e'''}}$, and ${\bf
Q_{SDW2}}=(0,0.495)$], and also two scattering wave vectors connecting the
$a_{1g}$-$e'_g$ FS pockets [${\bf Q_{ae'}}$ and ${\bf Q'_{SDW}}$]. At the same
time, these wave vectors also connect parts of the central $a_{1g}$ FS pocket
and the total magnetic susceptibility includes contribution from this
scattering too. In addition, there are two wave vectors, [${\bf Q_{AFM}}$ and
${\bf Q_{SDW1}}=(0,0.649)$] which arise due to the curved form of the central
$a_{1g}$ FS pocket. The pronounced peaks at all these wave vectors are present
in both the bare and the RPA magnetic susceptibility ($U$=0.15 eV). Again,
similar to the $a_{1g}$-band model, the magnetic response is not dominated by
the scattering at the commensurate wave vector ${\bf Q_{AFM}}$. The overall
picture of the magnetic response is consistent with the one presented in
Ref.~\onlinecite{Johannes2004}.
\begin{figure}
\includegraphics[angle=0,width=1.0\linewidth]{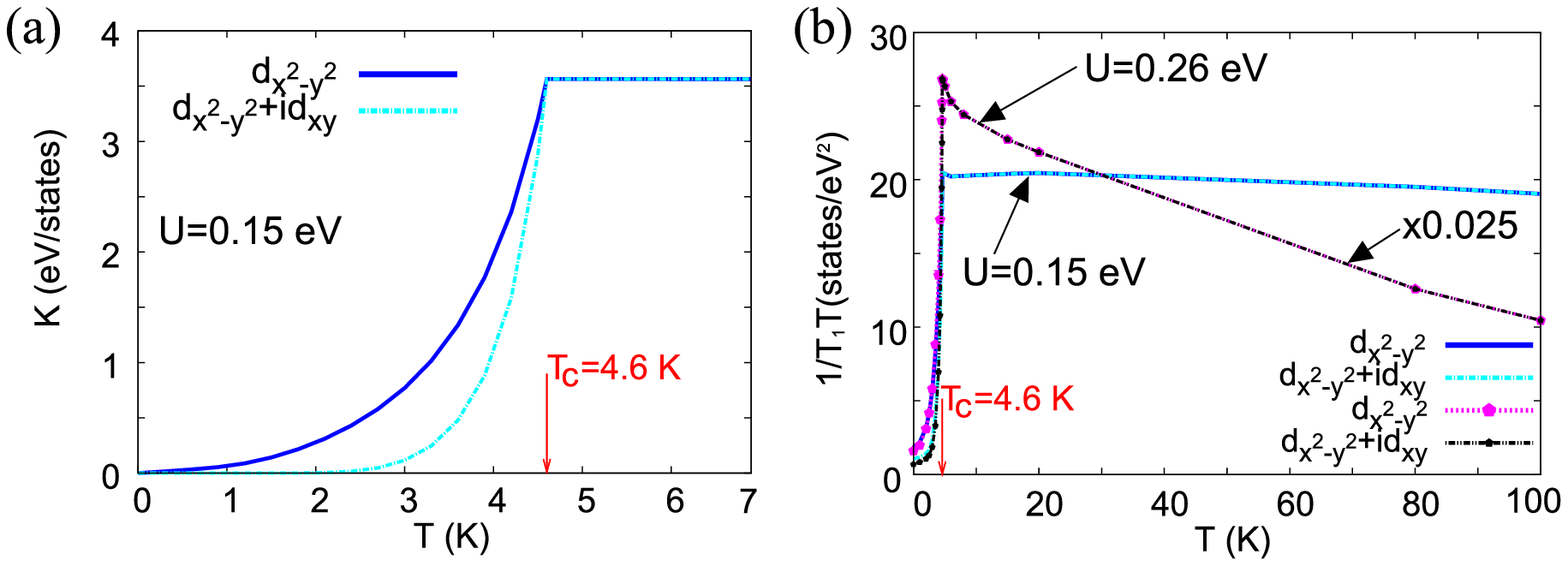}
\caption{Calculated  temperature dependence of the Knight shift $K(T)$ (a)
and the spin-lattice relaxation rate $1/T_1T$ (b) for the $t_{2g}$-band model.
Note, in (b) the curve for $U$=0.26 eV was scaled by a factor of 0.025.}
\label{fig:RigidBand_K_1_T1T}
\end{figure}

In the non-SC phase and the SC phase with $s$-wave order parameter
the behavior of $\chi({\bf q},\omega)$ at ${\bf q=Q_{AFM}}$ [see
Fig.~\ref{fig:rigidband_chi}(d)-(f)] is similar to the one in the
$a_{1g}$-band model. However, for the $d$-wave symmetry of the order
parameter, one finds that for $\omega \geq \Omega_g$ the states with
equal signs of the superconducting order parameter (second type of
the process ) contributes first, and the discontinuous jump in ${\rm
Im}[\chi_{0}({\bf Q_{AFM}},\omega)]$ occurs at higher energies. The
particular form of the FS in the realistic $t_{2g}$-band model and
more complicated band structure produce this effect. Therefore, the
resonance peak in ${\rm Im}[\chi_{RPA}({\bf Q_{AFM}},\omega)]$ may
in principle still exist, however, it occurs in a very narrow
interval of the $U$ values. This interval is determined by the
resonance condition in the superconducting state and by the
stability of a paramagnetic state above $T_c$. Here, we use
$U_{res}$=0.26 eV, which is more than twice smaller than in the
$a_{1g}$-band model.

Although the formation of the spin resonance is unrealistic for the
antiferromagnetic wave vector ${\bf Q_{AFM}}$ it may now occur at
other wave vectors. In Fig.~\ref{fig:rigidband_chi}(g)-(j) we
present the imaginary parts of $\chi_{0}({\bf q},\omega)$ and
$\chi_{RPA}({\bf q},\omega)$
at ${\bf Q_{SDW1}}$ and at ${\bf Q_{SDW2}}$. Here, one notices the
pronounced effects of the complicated $t_{2g}$-band structure at high energies
for the scattering at both wave vectors. Deviations from the linear-$\omega$
damping start already at low energies, smaller than $\Omega_g$. For
$U=U_{res}$ the spin-resonance is present at ${\bf Q_{SDW1}}$ for
both $d$-wave symmetries. However, at ${\bf Q_{SDW2}}$ the resonance
peak is present for $d_{x^2-y^2}+id_{xy}$-wave symmetry only.
Similar to the situation with ${\bf Q_{AFM}}$, this is due to smallness
of the allowed $U$ values.

\begin{figure}
\includegraphics[angle=0,width=1.0\linewidth]{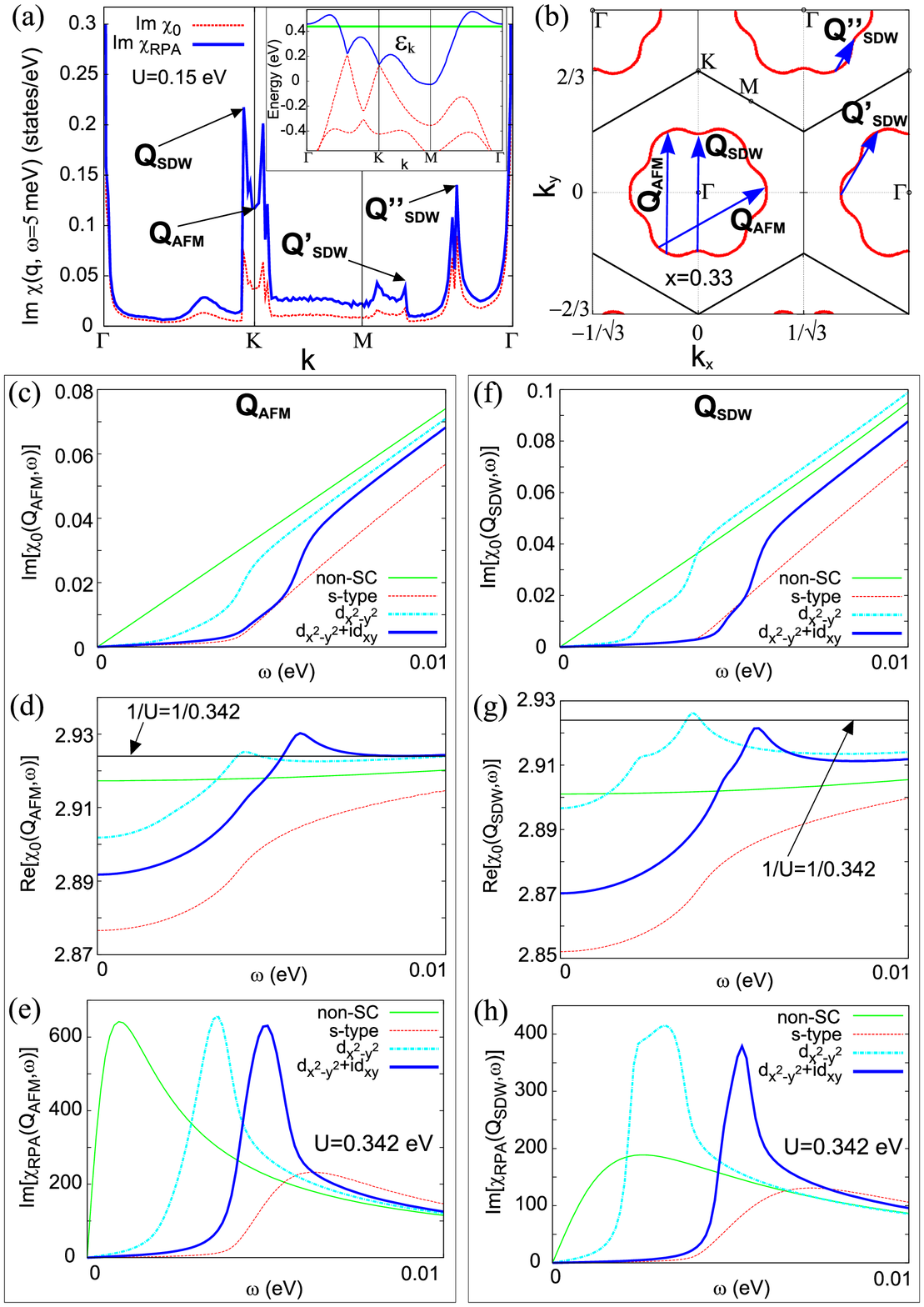}
\caption{ (Color online) Calculated results for the $t_{2g}$-band
model with enlarged crystal field splitting. (a) ${\bf q}$-dependence
of the ${\rm Im}[\chi_0({\bf q},\omega)]$ and the
${\rm Im}[\chi_{RPA}({\bf q},\omega)]$ at $\omega=5$ meV in the
normal (non-SC) phase. The scattering wave vectors ${\bf Q_{AFM}}$,
${\bf Q_{SDW}}$, ${\bf Q'_{SDW}}$, and ${\bf Q''_{SDW}}$  are
denoted by the arrows. The band dispersion is shown in the inset of
(a), where the bold (blue) curve denotes the topmost band used for
the susceptibility calculations, and the horizontal (green) line
stands for the chemical potential. (b) The calculated Fermi surface
with the corresponding scattering wave vectors. (c)-(e) The
calculated imaginary (c) and real (d) parts of the $\chi_{0}({\bf Q_{AFM}},\omega)$,
and the imaginary part of $\chi_{RPA}$ (e) in the normal
and in the SC state with various superconducting order parameters.
The same quantities are plotted in (f)-(h) at ${\bf q=Q_{SDW}}$.
Here we choose the amplitude of the superconducting order parameter
$\Delta_0=2$ meV. For the numerical purposes we also employ the
broadening of the Green's function, $\delta=0.2$ meV.}
\label{fig:rigidband100_chi}
\end{figure}
In Fig.~\ref{fig:RigidBand_K_1_T1T} we show the corresponding results for the
$1/T_1T$ and $K(T)$. Below superconducting transition temperature the behavior
is very similar to the results obtained for the simple $a_{1g}$-band model.
This is because below $T_c$ the symmetry of the superconducting gap and its
nodal structure determines the temperature dependencies of the $1/T_1T$ and the
$K(T)$ values. At the same time, notice the stronger AFM fluctuations in the
normal state. For almost the same value of $U$ as in
Fig.~\ref{fig:a1g_nn_K_1_T1T} this is due to the larger density of states at
the Fermi level (and the change of the Fermi velocity) than in the simple
$a_{1g}$-band model. Such a behavior is observed in the experimental NQR data
\cite{Zheng2006,Fujimoto2004,Ihara2005}. It is interesting to note that without
water the parent non-superconducting compound Na$_{0.33}$CoO$_2$ shows much
weaker AFM fluctuations \cite{Zheng2006}. In our theory the fluctuations occur
for the parent compound too. It probably demonstrates a possible significance
of the third dimension and, in particular, the bilayer splitting which may
reduce the two-dimensional AFM fluctuations in Na$_{0.33}$CoO$_2$.

Note, the presence of the $e'_g$ pockets on the FS can also lift the degeneracy
between the three $d$-wave states. Since in the $d_{x^2-y^2}$-wave SC state the
$e'_g$ FS pockets are fully gapped, the additional condensation energy is
gained [compare the topology of the line-nodes in Fig.~\ref{fig:a1g_SCgap}(a)
and FS topology in Fig.~\ref{fig:rigidband_chi}(b)]. For the $d_{xy}$-wave SC
state this gain in energy will be smaller [compare Fig.~\ref{fig:a1g_SCgap}(b)
and FS in Fig.~\ref{fig:rigidband_chi}(b)].

Presently, there is still a discussion on the details of the Fermi
surface topology in the water intercalated cobaltates. In
particular, ARPES experiments do not observe the $e'_g$-pockets at
the FS \cite{mzh2004, hby2004, shimojima2006}. It has been shown
that an inclusion of the electronic correlation within Gutzwiller
approximation may shift the $e'_g$-bands below the Fermi level
\cite{Zhou2005,Korshunov2007}, although this conclusion has been
challenged \cite{Ishida2005,Perroni2007,Liebsch2007}.
Another interpretation of this experimental result
relays on the disorder introduced by Na. As it was shown within LDA,
the scattering due to disorder can destroy the small $e'_g$-pockets
\cite{Singh2006}. For the superconducting polycrystalline samples,
recent experiments indicate that the oxonium ions, H$_3$O$^+$, may
introduce additional dopants \cite{Chen2004,Milne2004,Takada2004},
or result in oxygen vacancies reducing Co oxidation state
\cite{Barnes2005}. Though, this conclusion has been doubted by the
NMR experiments \cite{Mukhamedshin2005} which show the Co valence
state is insensitive to hydration and depends on the Na content
only. This was also confirmed later by the powder neutron
diffraction \cite{Viciu2006}.

In our study we further consider the $t_{2g}$-band model with increased crystal
filed splitting, $\delta\epsilon$=153 meV. This makes $e'_{g}$ band sink below
the Fermi level, as it is seen in the inset of
Fig.~\ref{fig:rigidband100_chi}(a). The behavior of the dynamical spin
susceptibility for $U$=0.15 eV at $\omega=5$ meV presented in
Fig.~\ref{fig:rigidband100_chi}(a) shows more similarity to the simple
$a_{1g}$-band model with additional features due to peculiarities (``rounded
hexagon'' form) of the large FS pocket as shown in
Fig.~\ref{fig:rigidband100_chi}(b). The scattering is most pronounced at the
wave vector ${\bf Q_{SDW}}=(0,0.633)$. There is also intensive scattering at
the wave vector ${\bf Q''_{SDW}}$, owing its appearance to the curved shape of
the FS.

Fig.~\ref{fig:rigidband100_chi}(c)-(e) and (f)-(h) displays the
magnetic susceptibility at ${\bf Q_{AFM}}$ and at ${\bf Q_{SDW}}$,
respectively. Contrary to both $a_{1g}$-band model and $t_{2g}$-band model
with $e'_g$ FS pockets, here we observe a well-defined linear
behavior of ${\rm Im}[\chi_0({\bf q},\omega)]$ in the considered
frequency range at these wave vectors. For the $d$-wave order
parameter, the behavior of the susceptibility resembles that in the
$t_{2g}$-band model with $e'_g$ FS pockets. Again one could find a narrow
range of parameters where the resonance peak exists, which we illustrate
in Fig.~\ref{fig:rigidband100_chi}(e),(h) for $U_{res}$=0.342 eV.
\begin{figure}
\includegraphics[angle=0,width=1.0\linewidth]{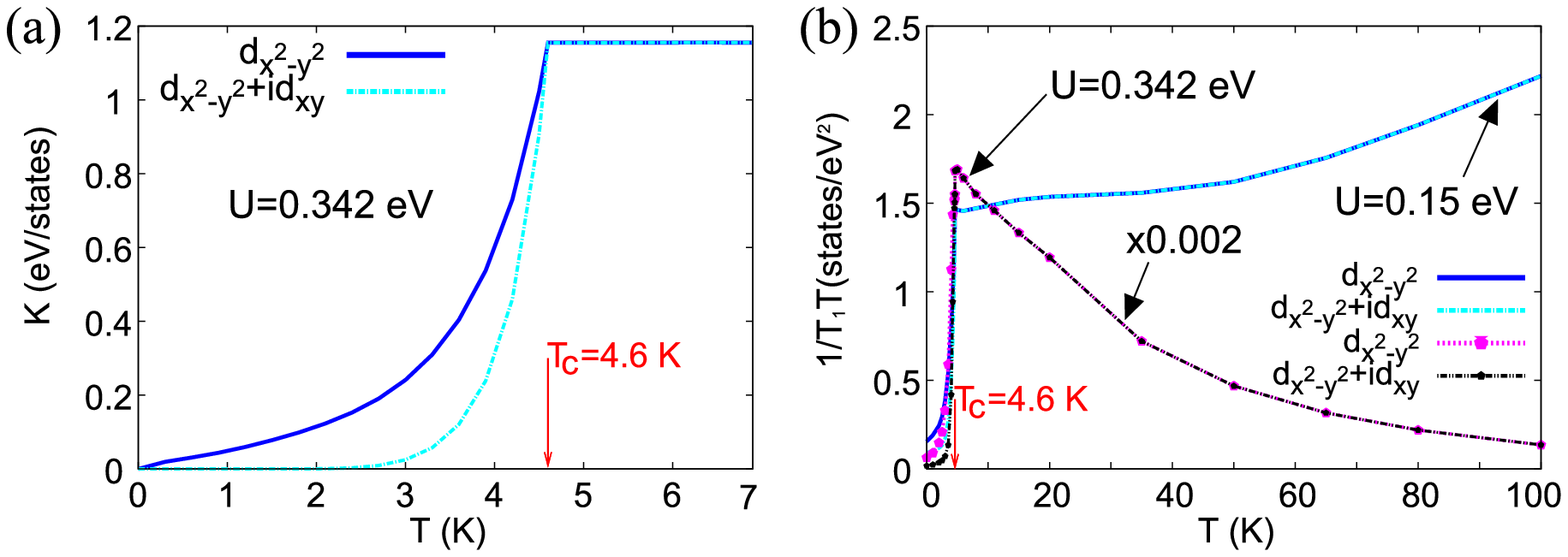}
\caption{(Color online) Calculated  temperature dependence of the
Knight shift $K(T)$ (a) and the spin-lattice relaxation rate
$1/T_1T$ (b) for the $t_{2g}$-band model without $e'_g$ FS pockets.
Note, in (b) the curve for $U$=0.342 eV was scaled by a factor of 0.002.}
\label{fig:RigidBand+100meV_K_1_T1T}
\end{figure}

Similarly, the change of the FS topology does not influence
significantly the temperature dependence of the Knight shift and the
spin-lattice relaxation rate above and below $T_c$. This is
illustrated in Fig.~\ref{fig:RigidBand+100meV_K_1_T1T} where we plot
both quantities as a function of temperature.

\section{Conclusion}

Our analysis of the dynamical spin susceptibility in application to
the Na$_x$CoO$_2$$\cdot y$H$_2$O have shown that the magnetic
response in the normal state is dominated by the incommensurate SDW
fluctuations at large momenta close to ${\bf Q_{AFM}}$. This is
consistent with experimental NQR data which shows a pronounced
AFM-like fluctuations in the temperature dependence of the
spin-lattice relaxation rate. It is interesting to note that the
presence of the $e'_g$-pockets at the Fermi surface is not affecting
significantly this result.
In the normal state we note the absence of ferromagnetic-like fluctuations.
This observation justifies our choice of spin-singlet order parameter, because
to induce the spin-triplet Cooper-pairing the fluctuations with small
momenta are required.
Below $T_c$ our results for $d_{x^2-y^2}$- or $d_{xy}$-wave (not shown)
symmetries of the superconducting order parameter are consistent with
experimental data which excludes nodeless $d_{x^2-y^2} + id_{xy}$-wave
symmetry. We further stress that the resonance peak, predicted previously
\cite{Li2004} for the simple $a_{1g}$-band model, is improbable for
the realistic band structure of Na$_x$CoO$_2$$\cdot y$H$_2$O.
Moreover, we find that even if present the resonance peak is
confined to the AFM wave vector and disappears away from it.

\begin{acknowledgments}
We would like to thank  A. Donkov, V. Yushankhai, A. Yaresko, and G.-q. Zheng
for useful discussions, and I. Mazin for useful comments. M.M.K. acknowledge
support form INTAS (YS Grant 05-109-4891) and RFBR (Grants 06-02-16100,
06-02-90537-BNTS). I.E. acknowledge support form Volkswagen Foundation.
\end{acknowledgments}

\end{document}